\documentclass[12pt,reprint,aps,groupedaddress,nofootinbib,prd,twocolumn]{revtex4-2}

\usepackage[english]{babel}
\usepackage[utf8]{inputenc}
\usepackage{amsmath}
\usepackage{mathbbol}
\usepackage{amssymb}
\usepackage{bbold}
\usepackage{graphicx,amsfonts}
\usepackage{epsfig}
\usepackage[colorlinks=true,
linkcolor=blue,
urlcolor=blue,
citecolor=blue]{hyperref}
\usepackage{bm}
\usepackage{mathrsfs}
\usepackage{enumerate}
\usepackage{amsthm}
\usepackage{bbm}
\usepackage{comment}
\usepackage{physics}
\usepackage{url}
\usepackage{upgreek}
\usepackage[svgnames]{xcolor}

\newcommand{\beq}{\begin{equation}}
\newcommand{\eeq}{\end{equation}}
\newcommand{\bea}{\begin{eqnarray}}
\newcommand{\eea}{\end{eqnarray}}
\newcommand{\bit}{\begin{itemize}}
\newcommand{\eit}{\end{itemize}}
\newcommand{\ben}{\begin{enumerate}}
\newcommand{\een}{\end{enumerate}}
\newcommand{\nn}{\nonumber}

\begin{document}

\title{Pseudospectrum of de Sitter black holes}

\author{Kyriakos Destounis$^{1,2}$, Valentin Boyanov$^{3}$, Rodrigo Panosso Macedo$^{4}$}
\affiliation{$^1$Dipartimento di Fisica, Sapienza Università di Roma, Piazzale Aldo Moro 5, 00185, Roma, Italy}
\affiliation{$^2$INFN, Sezione di Roma, Piazzale Aldo Moro 2, 00185, Roma, Italy}
\affiliation{${^3}$CENTRA, Departamento de F\'{\i}sica, Instituto Superior T\'ecnico -- IST, Universidade de Lisboa, Avenida Rovisco Pais 1, 1049 Lisboa, Portugal}
\affiliation{${^4}$ Niels Bohr International Academy, Niels Bohr Institute, Blegdamsvej 17, 2100 Copenhagen, Denmark}

\begin{abstract}
Pseudospectral analyses have broadened our understanding of ringdown waveforms from binary remnants, by providing insight into both the stability of their characteristic frequencies under environmental perturbations, as well as the underlying transient and non-modal phenomenology that a mode analysis may miss. In this work we present the pseudospectrum of scalar perturbations on spherically-symmetric black holes in de Sitter spacetimes. We expand upon previous analyses in this setting by calculating the pseudospectrum of Reissner-Nordstr\"om-de Sitter black holes, and revisit results regarding the stability of quasinormal modes under perturbations in several cases. Of particular note is the case of scalar quasinormal modes with angular parameter $\ell=0$, which possess a zero mode related to the presence of a cosmological horizon. We show that the non-trivial eigenfunction associated to this mode has a vanishing energy norm which poses a challenge in quantifying the magnitude of external perturbations to the wave equation's potential, as well as in calculating the pseudospectrum. Nonetheless, we present results which suggest that the spectral instability manifestation of $\ell=0$ scalar quasinormal modes is qualitatively the same as in other cases, in contrast to recent claims. We also analyze the stability of the fundamental mode for $\ell\ge1$, finding it to be spectrally stable, except for certain configurations in which a perturbation leads to a discontinuous overtaking of the fundamental unperturbed purely-imaginary mode by a perturbed complex quasinormal mode.
\end{abstract}

\maketitle

\section{Introduction}

One of the most important tests of general relativity (GR) in the strong-field regime is the detection of gravitational waves (GWs) from coalescing binary compact objects \cite{LIGOScientific:2021sio}. In particular, the ringdown phase of the remnant has a short burst during which it relaxes to stationarity through the emission of an exponentially damped oscillatory signal. This signal is characteristic to the remnant and corresponds to its resonant frequencies, i.e. its quasinormal modes (QNMs)~\cite{Kokkotas:1999bd,Nollert:1999ji,Berti:2009kk,Konoplya:2011qq}.

In practice, a GW measurement is analyzed by comparing it to templates generated from theoretical calculations in GR, or in other, modified theories of gravity. Such calculations are typically performed with certain simplifying assumptions, the most common one being that the coalescing objects form an isolated gravitational system. In other words, the astrophysical environment both at small scales (other gravitating objects in the vicinity of the binary) and at large scales (the cosmological background) are neglected. At first glance this assumption appears to be justified, as going beyond it does not seem to affect the dominant part of the resulting signals, i.e. the part resolved by current GW detectors. However, an analysis in the frequency domain reveals that the characteristics of the underlying system can change completely, altering the sub-dominant and late time aspects of the signals.

At large scales, it is known that changing the asymptotic structure of spacetime, where the system resides, changes some key features of the QNM spectrum. Particularly, in asymptotically flat spacetimes the discrete spectrum is accompanied by a continuous part corresponding to a branch cut in the Green's function of the evolution operator, which translates into a late-time inverse-polynomial decay of GWs~\cite{Leaver:1986gd,Gundlach:1993tp,Dafermos:2003yw,Hintz:2020roc,Dafermos:2021cbw}. On the other hand, if one includes a cosmological constant~\cite{Carroll:2000fy}, even the small one which is seemingly present in our universe~\cite{SupernovaSearchTeam:1998fmf,SupernovaCosmologyProject:1998vns,HST:2000azd}, the presence of a cosmological horizon changes the late-time decay to one which is exponentially damped~\cite{Brady:1996za,Molina:2003dc,Dafermos:2007jd,Hintz:2016gwb,Hintz:2016jak}, replacing the branch cut with a discrete set of non-oscillatory, purely damped modes.

At small scales, changes to the background of the system seem to affect the QNM spectrum even more significantly. It was first observed in Refs.~\cite{Aguirregabiria:1996,Nollert:1996rf,Nollert:1998ys} that slight small-scale changes to the effective potential in the wave equation which governs the evolution of perturbations leads to a disproportionately large displacement of the QNM frequencies in the complex plane. This then led to the seminal work on the spectral instability of QNMs in Schwarzschild black holes (BHs) \cite{Jaramillo:2020tuu}. A systematic analysis of the effect that perturbations to the effective potential have on the QNM spectrum was performed, accompanied by a calculation of the corresponding pseudospectrum~\cite{Trefethen:2005}. The instability was thus quantified by relating the migration of the QNM frequencies to the energy scale of the background perturbation as measured by the energy norm~\cite{Gasperin2021}. It was discovered that, among perturbations which preserve the asymptotic structure, those concentrated on small scales affect the spectrum the most, and that all overtones can be destabilized, but the fundamental mode seems to remain stable. This picture seems to be in tune with the BH spectroscopy program when the dominant QNM of GW ringdown signals is used for tests of GR. Nevertheless, higher overtones are important for a full BH spectral analysis~\cite{Giesler:2019uxc} and are therefore still sought after in GW ringdown data~\cite{London:2014cma,Carullo:2019flw,Isi:2021iql,Cotesta:2022pci,Isi:2022mhy,Finch:2022ynt,Isi:2023nif,Carullo:2023gtf,Isi:2019aib}. However, any precision tests of GR which make use of overtones need to be handled with care due to this spectral instability~\cite{Jaramillo:2021tmt,Berti:2022xfj} (see Ref. \cite{Destounis:2023ruj} for a review on BH spectroscopy and spectral instabilities). We also note that it has been found that even the fundamental QNM can be disproportionately destabilized \cite{Cheung:2021bol,Courty:2023rxk} when seemingly ``small" envinronmental bumps are added to the effective potential of the system at large distances~\cite{Barausse:2014tra,Cardoso:2021wlq,Cardoso:2022whc,Destounis:2022obl}, though the ``smallness" of these perturbations was not quantified in the same manner.

The overtone instability, as measured by the pseudospectrum in the energy norm, was subsequently analyzed in a variety of spacetimes with different asymptotic structures and boundaries, such as Reissner-Nordstr\"om (RN) \cite{Destounis:2021lum}, Schwarzschild-de Sitter (SdS) \cite{Sarkar:2023rhp}, Schwarzschild-anti de Sitter black branes \cite{Arean:2023ejh} and BHs \cite{Cownden:2023dam,Boyanov:2023}, as well as horizonless exotic compact objects with a reflective surface~\cite{Boyanov:2022ark}. Each of these spacetimes, even before being perturbed, have an overall QNM spectrum which is quite different from the Schwarzschild case alone. In all these cases, the qualitative picture of spectral instability is similar, though there are certain differences. For instance, in the exotic compact object case, the stability extends past the fundamental mode and includes the first overtones, when the object is sufficiently compact. Another series of differences were observed in the SdS case, which we aim to revisit in the present work, also extending part of the analysis to the Reissner-Nordström-de Sitter (RNdS) family of BH solutions.

The RNdS case is of particular importance, given the richness of its spectral content. Particularly, its QNMs are typically categorized into three distinct families. Firstly, there is the family of \emph{photon sphere} (PS) QNMs, associated to the angular frequency and instability timescale of null geodesics at the PS \cite{Cardoso:2008bp,Konoplya:2017wot}. Second, there is a set of purely imaginary modes (non-oscillatory damped modes) associated to the presence of the cosmological horizon, referred to as \emph{de Sitter} (dS) modes \cite{Lopez-Ortega:2006aal,Jansen:2017oag,Cardoso:2017soq,Destounis:2019hca,Konoplya:2022xid}. Third, there is another family of purely damped modes, for which the decay timescales are typically very short (their imaginary parts are very large), except when the BH is near extremality, i.e. when its charge is large enough for its inner and outer horizons to be close to each other; this family is referred to as the \emph{near-extremal} (NE) family of modes \cite{Richartz:2014jla,Hod:2017gvn,Cardoso:2017soq,Cardoso:2018nvb,Destounis:2018qnb,Destounis:2019hca,Destounis:2019omd}. The different behaviors of each of these families for different parameters of the RNdS BH present an ideal testing ground for spectral and pseudospectral analyses, since they represent different ``large scale" settings and thus allow for a more thorough analysis of environmental effects on the spectrum and its (in)stability in different environments. Another important characteristic is the lack of the aforementioned branch cut in the Green's function. When present, the branch cut manifests itself as randomly distributed and non-convergent eigenvalues on the imaginary axis within our numerical scheme~\cite{Jaramillo:2020tuu}. Its absence provides us with a pristine setup to study the stability of the QNMs along the imaginary axis.

The goal of the present work is to revisit and expand upon previous work done on the spectral instability of SdS and RNdS spacetimes~\cite{Sarkar:2023rhp}. We begin by analyzing the case of $\ell=0$ scalar QNMs, where a surprisingly strong spectral stability was reported in \cite{Sarkar:2023rhp} for both SdS and RNdS geometries. We show that this apparent stability is likely just an artefact of an issue with the norm used to measure the size of perturbations. Particularly, the energy norm of the zero-mode contained in the $\ell=0$ spectrum vanishes, implying that the operator associated with the norm has a zero eigenvalue and thus its inverse is ill-defined. As this inverse is used in calculations of the norms of perturbations, the issue can lead to numerically non-convergent measures of their size, and attempts to normalize them would in turn lead to practically vanishing values. We show that under appropriately scaled perturbations, overtones do indeed seem to be unstable. Additionally, we find that despite the strict non-convergence of the norm, at a given finite numerical resolution there are no divergences. The choice of a finite number of gridpoints therefore serves as cutoff to the divergence, and can provide a qualitative (but not quantitative) example of a well behaved norm. We find that in such a norm, the pseudospectrum also suggests that the standard spectral instability is present for $\ell=0$, though it does not serve to quantify it.\footnote{It is worth noting that the pseudospectrum is in fact not used to directly quantify the BH QNM instability, neither in this nor in previous works, due to convergence issues which are more general in nature, as discussed in~\cite{Boyanov:2023}.}

We then calculate the pseudospectra of scalar QNMs for various SdS and RNdS BHs, for choices in the parameter space where different families of modes are dominant, and for angular numbers $\ell\geq 1$, where the energy norm is well behaved. Combined with calculations of the spectrum with particular perturbations to the effective potential, we arrive at a picture of spectral instability that is qualitatively similar in almost all cases. We find that overtones are almost always unstable under relatively small perturbations, and we quantify the stability of the fundamental mode. We find that the mode which is initially fundamental, whether it be a PS mode or a purely imaginary one, is by itself spectrally stable under perturbations of energy norm up to order $10^{-1}$ (that is, its relative displacement is less than the size of the perturbation), as hinted at in \cite{Sarkar:2023rhp}. However, when the first overtone has an imaginary part which is close enough to that of the fundamental mode that the difference lies within the range of displacement by a small perturbation, which can occur for particular values in the parameter space of dS BHs, an \emph{overtaking instability} occurs, whereby the displacement of the overtone can bring it below the the (previously) fundamental mode, making it the new longest-lived one. Despite being dubbed an instability, we emphasize that it can and does occur when both modes involved are individually stable, i.e. when after the perturbation they each have a relative displacement less than the magnitude of the perturbation. This brings us to the additional observation that the only overtones which appear to be spectrally stable are the ones which have an imaginary part very close to that of the (initially) fundamental mode. We also note that the phenomenon of overtaking was also observed in Ref.~\cite{Cheung:2021bol}, where the effective potential was perturbed in a different manner.

\section{Asymptotically de-Sitter black holes}
\subsection{Schwarzschild coordinates}

In the standard Schwarzschild coordinates $(t,r,\theta,\varphi)$, the spherically-symmetric electrovacuum solution of the Einstein-Maxwell equations with a positive cosmological constant, i.e. the RNdS geometry \cite{Griffiths:2009dfa}, has a line element of the form
\begin{equation}\label{line_element_SdS}
    ds^2=-f(r)dt^2+f(r)^{-1}dr^2+r^2 d\Omega^2,
\end{equation}
where $d\Omega^2=d\theta^2 + \text{sin}^2\theta\,d\varphi^2$ is the line element of the unit 2-sphere and the redshift function $f(r)$ reads
\beq
\label{eq:func_MQLambda}
f(r)=1-\frac{2M}{r} + \dfrac{Q^2}{r^2}-\frac{\Lambda r^2}{3}.\\
\eeq
Here, $M$ and $Q$ are the BH mass and electric charge, respectively, and $\Lambda>0$ is the cosmological constant. By setting $Q=0$ we obtain the SdS spacetime \cite{Griffiths:2009dfa}. With the redshift function we define the tortoise coordinate $r_*$ such that $f(r)=dr/dr_*$. Sub-extremal RNdS geometries possess three horizons, which are located at the positive roots of $f(r)$. We will use $r=r_\Lambda$ to designate the cosmological horizon, $r=r_h$ the event horizon and $r=r_c$ the Cauchy horizon radius. For a given value of the parameters $M$, $Q$ and $\Lambda$, these horizons are uniquely determined by the relations
\bea
M &=& \dfrac{(r_h + r_\Lambda)(r_h + r_c)(r_c + r_\Lambda)}{2 ( r_h^2 + r_h r_\Lambda  + r_h r_c  + \ r_\Lambda^2 +  r_c^2 +   r_c  r_\Lambda ) }, \label{eq:M} \\
Q^2 &=& \dfrac{r_h r_\Lambda r_c (r_h + r_\Lambda + r_c)}{r_h^2 + r_h r_\Lambda  + r_h r_c  + \ r_\Lambda^2 +  r_c^2 +   r_c  r_\Lambda }, \label{eq:Q}\\
\Lambda &=& \dfrac{3}{  r_h^2 + r_h r_\Lambda  + r_h r_c  + \ r_\Lambda^2 +  r_c^2 +   r_c  r_\Lambda   }. \label{eq:Lambda}
\eea
The redshift function $f(r)$ has one additional, negative root $r_o = -(r_h + r_c + r_\Lambda)$, which lies outside the range of $r$ corresponding to the spacetime description. Even though the most standard representation of RNdS is given by Eq. \eqref{eq:func_MQLambda}, in what follows it is useful to express it in its factorized form, with respect to the horizons and the negative root, as
\beq
f(r)= -r^2 \dfrac{\Lambda}{3}\left(1-\dfrac{r_h}{r}\right)\left(1-\dfrac{r_c}{r}\right)\left(1-\dfrac{r_\Lambda}{r}\right)\left(1-\dfrac{r_o}{r}\right).
\eeq
This form is convenient for constructing the hyperboloidal slices~\cite{PanossoMacedo:2023qzp}, which we will do in Sec. \ref{sec.HypSlices}. Furthermore, it also makes it easier to identify the extremal regimes: extremal charge corresponds to $r_h=r_c$, and extremal cosmological constant (i.e. the Nariai limit) to $r_h=r_\Lambda$.

Parametric studies on RNdS spacetimes may, therefore, follow two strategies: (i) one fixes the parameters $M$, $Q$ and $\Lambda$ and them determines the horizons $r_h$, $r_c$ and $r_\Lambda$ by solving Eqs.~\eqref{eq:M}-\eqref{eq:Lambda}, or (ii) one fixes the horizons $r_h$, $r_c$ and $r_\Lambda$ and then calculates $M$, $Q$ and $\Lambda$ directly from Eqs.~\eqref{eq:M}-\eqref{eq:Lambda}. The former is more practical when expressing dimensional observables with respect to a typical length scale such as the BH mass $M$. This is the more common approach in the literature (see e.g. \cite{Cardoso:2017soq,Cardoso:2018nvb,Destounis:2018qnb,Liu:2019lon,Destounis:2019hca,Destounis:2019omd,Mascher:2022pku,Courty:2023rxk}), which allows one to prescribe dimensionless values such as $Q/M$ and $\Lambda M^2$. The expressions for the hyperboloidal framework from Sec.~\ref{sec.HypSlices}, on the other hand, assume simpler forms if one employs the horizon $r_h$ as the typical length scale. For that purpose, strategy (ii) is more convenient. Thus, while we follow the typical convention of setting $Q/M$ and $\Lambda M^2$ when choosing the parameters for the examples we use in Sec.~\ref{sec:PseudoSpectra}, in the numerical results the QNM frequency space is represented in units of $r_h$ as a direct consequence of the hyperboloidal framework which we will introduce.

\subsection{Hyperboloidal framework}\label{sec.HypSlices}

Following Ref.~\cite{Jaramillo:2020tuu}, together with the techniques reviewed in Ref.~\cite{PanossoMacedo:2023qzp}, we introduce a hyperboloidal coordinate system that imposes geometrically the boundary conditions associated to waves propagating into the BH and cosmological horizons \cite{Zenginoglu:2011jz}. Specifically, the scri-fixing approach \cite{Zenginoglu:2007jw}, with definitions from Ref.~\cite{PanossoMacedo:2023qzp} introduces hyperboloidal coordinates $(\tau, \chi, \theta, \varphi)$ via
\begin{equation}\label{hypercoords}
t=\lambda \bigg( \tau - H(\chi) \bigg),\, \quad r=\dfrac{r_h}{\chi}.
%
\end{equation}
Here, $\lambda$ is some characteristic length scale of spacetime, and the height function $H(\chi)$ is constructed by requiring that $\tau=$ constant slices are spacelike hypersurfaces penetrating the future event and cosmological horizons.

Following previous works~\cite{Ansorg:2016ztf,PanossoMacedo:2018hab,PanossoMacedo:2019npm,PanossoMacedo:2023qzp}, we will choose a height function corresponding to the so-called \emph{minimal gauge}. The particular form of the radial compactification already reflects a first step associated to this choice. In terms of the compact radial coordinate $\chi$, the horizons are located at $\chi_h = 1$ (event horizon), $\chi_c = r_h/r_c$ (Cauchy horizon) and $\chi_\Lambda = r_h/r_\Lambda$ (cosmological horizon). We also identify the negative root of $f(r)$ as $\chi_o = r_h/r_o$. Due to the inversion in the radial transformation, the horizons now satisfy the relation $\chi_\Lambda \leq \chi_h \leq \chi_c$. We are interested in the behavior of scalar fields propagating in the external region of a sub-extremal BH, delimited by the domain of the observable universe, i.e. $\chi \in [\chi_\Lambda, \chi_h]$. 

To fix $H(\chi)$ in the minimal gauge, it is convenient to first define the dimensionless tortoise coordinate~\cite{PanossoMacedo:2023qzp}
\bea
x(\chi) &=& \dfrac{r_*(r(\chi))}{\lambda} \\
&=& \underbrace{ x_h(\chi) + x_c(\chi) + x_\Lambda(\chi)}_{x_{\rm sing}(\chi)} + x_{\rm reg}(\chi),
\label{eq:x}
\eea
with 
\bea
x_{\rm h}(\chi) &=&- \dfrac{3\, \chi_h}{\lambda r_h \Lambda}\dfrac{\ln(\chi_h-\chi)}{(1 - \chi_h/\chi_{c})(1 - \chi_h/\chi_\Lambda)(1 - \chi_h/\chi_o)},  \nn \\
\\
x_{\rm c}(\chi) &=&- \dfrac{3 \, \chi_{c}}{\lambda r_h \Lambda}\dfrac{\ln(\chi_{c}-\chi)}{(1 - \chi_{c}/\chi_h)(1 - \chi_{c}/\chi_\Lambda)(1 - \chi_{c}/\chi_o)}, \nn \\
\\
x_{\Lambda}(\chi) &=& - \dfrac{3 \, \chi_\Lambda}{\lambda r_h \Lambda}\dfrac{\ln(\chi-\chi_{\Lambda})}{(1 - \chi_{\Lambda}/\chi_h)(1 -  \chi_{\Lambda}/\chi_{\rm C})(1 -  \chi_{\Lambda}/\chi_o)}, \nn \\
\\
x_{\rm reg}(\chi)&=& - \dfrac{3 \chi_o}{\lambda r_h \Lambda}\dfrac{\ln(\chi-\chi_{o})}{(1 - \chi_{o}/\chi_h)(1 -  \chi_{o}/\chi_{c})(1 -  \chi_o/\chi_{\Lambda})}. \nn \\
\eea
As pointed out in Eq.~\eqref{eq:x}, the terms $x_{\rm h}(\chi)$, $x_{\rm c}(\chi)$ and $x_{\Lambda}(\chi) $ contribute to the singular piece of $x(\chi)$ because they diverge at $\chi_h$, $\chi_c$ and $\chi_\Lambda$, respectively. The regular piece $x_{\rm reg}(\chi)$, resulting from the negative root $\chi_o$, assumes a finite value in the entire domain, including all horizons. 

The height function is constructed straightforwardly \cite{PanossoMacedo:2023qzp} by simply changing the signs of the $x_{\Lambda}(\chi)$ and $x_{\rm reg}(\chi)$, i.e.
\beq
H(\chi) = x_h(\chi) + x_c(\chi) - x_\Lambda(\chi) - x_{\rm reg}(\chi).
\eeq
This practical approach is a direct consequence of the out-in strategy for constructing hyperboloidal slices in the minimal gauge, and it ensures that the surfaces $\tau =$ constant remain spacelike in the domain $\chi \in [\chi_\Lambda, \chi_h]$, for all parameters $Q/M$ and $\Lambda M^2$ \cite{PanossoMacedo:2023qzp}.

With the coordinate transformation \eqref{hypercoords}, the line element \eqref{line_element_SdS} conformally rescales as $ds^2 = \Xi^2 d\bar s^2$, with $\Xi = \sigma/\lambda$ and \cite{PanossoMacedo:2023qzp}
\beq
d\bar s^2 = - p(\chi) d\tau^2 + 2 \gamma(\chi) d\tau d\chi + w(\chi) d\chi^2 + \dfrac{r_h^2}{\lambda^2}d\Omega^2.
\eeq 
The metric functions relate to $x(\chi)$ and $H(\chi)$ via
\begin{eqnarray}
\label{eq:funcs_wave}
p =- \dfrac{1}{x'}, \quad \gamma = p H', \quad w = \dfrac{1 - \gamma^2}{p}.
\end{eqnarray}
In particular, $p(\chi)$ assumes the form
\beq
\label{eq:func_p}
p(\chi) = \dfrac{r_h^2 \Lambda}{3}\left( 1- \chi \right) \left( 1 - \dfrac{\chi}{\chi_c }  \right) \left(  \dfrac{\chi}{\chi_\Lambda} -1 \right)\left( 1 -   \dfrac{\chi}{\chi_o}\right),
\eeq
which will be explicitly used in Sec.~\ref{sec:zero mode}. The expressions for $\gamma(\chi)$ and $w(\chi)$ are quite lengthy and thus we will omit them here, though their explicit representations can be easily obtained.

\subsection{Scalar perturbations and quasinormal modes}

At linear order, the propagation of scalar perturbations $\Psi$ is governed by the free Klein-Gordon equation
\begin{equation}
    \Box \Psi=\frac{1}{\sqrt{-g}}\left(g^{\mu\nu}\sqrt{-g}\Psi_{,\nu}\right)_{,\mu}=0,
\end{equation}
where $g$ is the determinant of the metric tensor of the background geometry $g_{\mu\nu}$, and a comma in the subscript denotes partial differentiation. A standard separation of variables of the form
\begin{equation}
\Psi(t,r,\theta,\varphi)=\sum_{\ell,m}\frac{\psi_\ell(t,r)}{r}Y_{\ell,m}(\theta,\varphi),
\end{equation}
where $Y_{\ell,m}$ are the spherical harmonics, leads to the wave equation
\begin{equation}\label{waveeq}
-\psi_{,tt}+\psi_{,r_*r_*}-V\psi=0.
\end{equation}
The effective potential $V$ for scalar perturbations has the form~\cite{Berti:2009kk}
\begin{align}\label{Vsc}
V=f(r)\left(\frac{\ell(\ell+1)}{r^2}+\frac{f'(r)}{r}\right),
\end{align}
where prime denotes differentiation with respect to the function's variable. While gravitational and electromagnetic perturbations can also be reduced to solving an equation of the type \eqref{waveeq}, we find no qualitative difference in the results presented below between these different types of perturbations, and we thus limit the discussion to the scalar case, for which the spectrum is already quite intricate \cite{Cardoso:2017soq} and leads to interesting phenomenology \cite{Sarkar:2023rhp}.

The region of interest for our analysis, as discussed, lies in the static patch between the event horizon $r_{\rm h}$ and the cosmological horizon $r_{\Lambda}$. This translates to solutions in the static region that satisfy a purely ingoing boundary condition at the event horizon,
\begin{equation}\label{ingoing}
\left(\psi_{,t}-\psi_{,r_*}\right)|_{r=r_{\rm h}}=0,
\end{equation}
and a purely outgoing boundary condition at the cosmological horizon,
\begin{equation}\label{outgoing}
\left(\psi_{,t}+\psi_{,r_*}\right)|_{r=r_{\rm c}}=0.
\end{equation}
The vanishing of the effective potential $V$ at the horizons makes these conditions compatible with the wave equation. The QNMs are then obtained by considering a Fourier decomposition of the form $\psi(t,r_*)\sim\psi(r_*) e^{i\omega t}$ and solving \eqref{waveeq}, along with the aforementioned boundary conditions, as an eigenvalue problem. However, the boundary value problem, as cast in the present coordinate system, becomes singular at the horizons. As in previous works, we resolve this issue geometrically by switching to the hyperboloidal slicing described in the previous subsection.

\subsection{Non-modal tools: pseudospectrum, energy norm and the eigenvalue condition number}

We follow Ref.~\cite{Jaramillo:2020tuu} and cast the QNM eigenvalue problem in the hyperboloidal frame in order to calculate the spectrum and pseudospectrum. The first step consists of mapping the wave equation in $(t,r_*)$ coordinates to its equivalent form in the hyperboloidal $(\tau,\sigma)$ coordinates. For the sake of completeness we rewrite the relevant expressions here (see Ref.~\cite{PanossoMacedo:2023qzp} for a systematic review). With a reduction of order in time $\phi\equiv\psi_{,\tau}$, Eq.~\eqref{waveeq} can be rewritten as a system of two equations,
\begin{equation}\label{matrix_evolution}
	u_{,\tau}=i L u, \quad 	L =\frac{1}{i}\!
	\left(
	\begin{array}{c  c}
		0 & 1 \\
		L_1 & L_2
	\end{array}
	\right), \quad u=\left(
	\begin{array}{c}\psi \\ \phi \end{array}\right),
\end{equation}
where we have defined the differential operators
\begin{equation}
\label{eq:Operators_L1_L2}
	\begin{split}
		L_1&=\frac{1}{w(\chi)}\bigg[\partial_\chi \left(p(\chi) \partial_\chi \right) - q(\chi) \bigg],\\
		L_2&=\frac{1}{w(\chi)}\bigg[2\gamma(\chi)\partial_\chi + \partial_\chi\gamma(\chi)\bigg],
	\end{split}
\end{equation}
with the functions $p(\chi)$, $\gamma(\chi)$ and $w(\chi)$ given in Eq.~\eqref{eq:funcs_wave}. In the hyperboloidal framework, the potential rescales to
\beq
\label{eq:hyp_pot}
q = \lambda^2 \dfrac{V}{p}.
\eeq
The discrete set of QNMs, $\omega_n$ with $n$ being the overtone number, comprise the spectrum $\sigma(L)$ of the differential operator $L$ in \eqref{matrix_evolution}. Expanding this notion, the $\epsilon$-\emph{pseudospectrum} $\sigma^{(\epsilon)}(L)$ is defined as
\begin{equation}\label{pseudospectrum}
\sigma^{(\epsilon)}(L)=\{\omega\in\mathbb{C}:\|R_L(\omega)\|>1/\epsilon\},
\end{equation}
where $R_L(\omega)=(L-\omega\mathbb{I})^{-1}$ is the \emph{resolvent operator}. In the $\epsilon\to 0$ limit, the set of points $\omega$ reduces to the spectrum $\omega_n$, while for larger values of $\epsilon$ the boundary of the set corresponds to points in the complex plane which are ``further away" from being eigenvalues. The quantity $\epsilon$ provides a measure of this ``closeness" of points to the spectrum, which has a clear interpretation in terms of perturbations to the underlying operator. If one assumes a perturbation 
\begin{equation}\label{perturbed operator}
L\to L+ \epsilon \delta L,
\end{equation}
where $\|\delta L\|=1$ and $\epsilon>0$, then the spectrum can migrate anywhere inside the $\epsilon$-pseudospectral region of the same $\epsilon$. Considering all possible perturbations in fact provides an alternative definition of the pseudospectrum \cite{Trefethen:2005}. The shape and size of the $\epsilon$-pseudospectral regions therefore quantifies the spectral stability of $L$. Particularly, if their boundaries form concentric circles of radius $\epsilon$, then the operator is spectrally stable, while other behaviors are indicative of instability, as seen, e.g. in Refs.~\cite{Jaramillo:2020tuu,Destounis:2021lum}.

We note that the pseudospectrum depends on the choice of norm with which one computes $\|R_L(\omega)\|$, and $\|\delta L\|$. As in previous works~\cite{Jaramillo:2020tuu, Destounis:2021lum,Boyanov:2022ark}, we will use the \emph{energy norm} \cite{Gasperin2021}, which provides a physical measure of the magnitude of solutions of the wave equation, and by extension of operators which can act on them. The definition of this norm for a spherically-symmetric BH in dS spacetime, for a solution vector $u$ is given by
\begin{equation}\label{energy norm}
\|u\|^2_{_E}=\frac{1}{2}\int_{\chi_{{\rm \Lambda}}}^{\chi_{{\rm h}}} \left(w(\chi)|\phi|^2
+ p(\chi)|\partial_\chi\psi|^2 + q(\chi) |\psi|^2\right) d\chi,
\end{equation}
which allows one to introduce the scalar product between two vectors $u$ and $v$ via
\begin{eqnarray}
    \label{energy norm scalar}
    \displaystyle
& \displaystyle |\langle u,v\rangle|_{_E}=\frac{1}{2}\int_{\chi_{{\rm \Lambda}}}^{\chi_{{\rm h}}} \Big(w(\chi) \phi_v^* \phi_u  \notag \\
&  + p(\chi) \partial_\chi\psi_u^* \, \partial_\chi\psi_v  + q(\chi) \psi_u^* \,\psi_v \Big) d\chi.
\end{eqnarray}
The pseudospectra of the $L$ operator which we calculate below are non-trivial, that is, their contours do not form concentric circles around the spectrum. This is a consequence of the non-selfadjoint (more generally, non-normal) nature of $L$ in this norm. Another measure of this feature is the the \emph{eigenvalue condition number} of the spectrum~\cite{Trefethen:2005,Jaramillo:2020tuu}. For the QNM operator $L$ and its eigenvalues $\omega_n$, the corresponding left and right eigenvectors $u_n$ and $v_n$, respectively, are characterized by the following eigenvalue expressions
\begin{equation}
    L^\dagger\, u_n=\Tilde{\omega}_n\, u_n, \qquad L\, v_n=\omega_n\, v_n,
\end{equation}
where $\Tilde{\omega}_n$ is the complex conjugate of $\omega_n$. If we perturb the operator, as in Eq. \eqref{perturbed operator}, then the perturbed eigenvalues $\omega_n(\epsilon)$ satisfy \cite{Jaramillo:2020tuu}
\begin{equation}
    |\omega_n(\epsilon)-\omega_n|\leq \epsilon\, \kappa_n,
\end{equation}
with the eigenvalue condition number being defined as
\begin{equation}\label{condnumb}
    \kappa(\omega_n)=\kappa_n\equiv \frac{\|u_n\|_{_E}\,\|v_n\|_{_E}}{|\langle u_n,v_n\rangle|_{_E}}.
\end{equation} 
When the operator $L$ is normal, i.e. when $[L^\dagger,L]=0$, the left and right eigenvectors can be diagonalized in the same basis. Therefore, from Eq. \eqref{condnumb}, $\kappa_n=1$ for normal operators, which is indicative of spectral stability. In these cases, generic perturbations of order $\epsilon$ to the operator can only lead to migration of $\omega_n(\epsilon)$ up to a distance $\epsilon$ in the complex plane. In contrast, when $L$ is non-selfadjoint, then the left and right eigenvectors are not necessarily collinear and can even become almost orthogonal, which leads to a very large condition number, as can be seen from Eq.~\eqref{condnumb}; a minuscule perturbation $\epsilon$ to $L$ can then lead to a disproportionately large migration of $\omega_n(\epsilon)$ in the complex plane. Therefore, $\kappa_n$ provides another measure of the sensitivity of eigenvalues to perturbations, once again associated with our choice of norm.

\subsection{Numerical discretization}\label{sec:NumDiscretization}

As in Ref.~\cite{Jaramillo:2020tuu}, the calculations presented below have been performed with a Chebyshev spectral discretization method with $N$ grid points along the compactified radial coordinate $\chi$. In particular, Appendix C from Ref.~\cite{Jaramillo:2020tuu} shows that the energy scalar product has a Chebyshev representation in terms of the discrete vector $\ket{u}$ and the energy operator $\hat H$ as
\begin{equation}\label{matrix evolution}
	|\langle u,v\rangle|_{_E} = \bra{u} \hat H \ket{u}.
\end{equation}
Then, the discrete representation of the adjoint operator  $\hat L^\dagger$ in the the energy norm follows from the discrete energy operator $\hat H$ via
\begin{equation}\label{dagger operator}
\hat L^\dagger = \hat H^{-1} \cdot \left(\hat  L^t \right)^* \cdot \hat H.
\end{equation}

From the energy scalar product, one can also derive~\cite{Jaramillo:2020tuu} an induced norm $\| \hat A \|_{_E}$ for a given matrix $\hat A$ describing a discrete representation of some physical quantity via
\begin{equation}
\| \hat A \|_{_E} = \sqrt{ \rho\left( A^\dagger A \right) }, \quad \rho\left( M \right) = \max_{m \in \sigma(M)} \left\{ m \right\},
\end{equation}
where the adjoint $A^\dagger$ is constructed through Eq.~\eqref{dagger operator}.

\section{Pseudospectrum of de Sitter black holes}\label{sec:PseudoSpectra}

In a recent work~\cite{Sarkar:2023rhp}, the spectral stability of SdS and RNdS BHs was studied. For SdS, the QNM spectrum of perturbed versions of the evolution operator $L$ was calculated, along with a topographical map of the pseudospectrum of some particular cases, using the numerical implementation of the definition~\eqref{pseudospectrum} discussed above. For RNdS, only the perturbed QNM spectra were investigated, while no pseudospectra calculations were performed. In the present work, we aim to corroborate, reinterpret, and expand upon some of these results. We focus on the case of a scalar field, as a representative example which covers all behaviors of interest, as we will see in the following. We split our presentation into two subsections covering the SdS and RNdS cases, though some results will be common to both.

\subsection{Schwarzschild-de Sitter black holes}
\subsubsection{The special case of $\ell=0$ scalar perturbations}\label{sec:zero mode}

We begin by reinterpreting one particular result of Ref.~\cite{Sarkar:2023rhp}: the apparent spectral stability of the $\ell=0$ scalar QNMs. No migration of scalar QNMs with $\ell=0$ was observed in that work, even when the effective potential was perturbed with (seemingly) high-amplitude fluctuations. In what follows we identify the particular features for the $\ell=0$ case which led to this conclusion. We will focus on SdS geometries, though we note that the results are independent of the BH charge and thus generalize to the broader family of RNdS.

Linear-order scalar perturbations in dS spacetimes with angular number $\ell = 0$ have one particular feature which distinguished them from other types of perturbations and other angular numbers. The eigenvalue problem related to Eq.~\eqref{matrix evolution} in the frequency domain has a non-trivial zero mode, i.e. the problem has a non-vanishing eigenvector $\phi_0(\chi)$ associated with the zero eigenvalue $\omega = 0$, that is regular at the event and cosmological horizon. For the zero-mode function $\phi_0(\chi)$, Eq. \eqref{matrix_evolution} becomes
\begin{equation}
    \label{zero_mode equation}
    \dfrac{d}{d\chi}\bigg( p(\chi)\, \phi_0'(\chi) \bigg) - q_0(\chi) =0,
\end{equation}
with $p(\chi)$ given by Eq.~\eqref{eq:func_p} and $q_0(\chi)$ given by setting $\ell=0$ in Eqs.~\eqref{Vsc} and \eqref{eq:hyp_pot}, i.e.
\beq
    q_0(\chi) = \dfrac{r_h^2 \Lambda}{3 \chi^2}\left[ - 2 + \chi^3 \left(1 + \dfrac{r_\Lambda}{r_h} \right) \dfrac{r_\Lambda}{r_h} \right].
\eeq
A solution to Eq.~\eqref{zero_mode equation}, regular at $r=r_h$ and $r=r_\Lambda$, is
\begin{equation}
    \phi_0(\chi) = \dfrac{1}{\chi}.
\end{equation}
The zero mode, however, does not contribute to the energy of the system. Indeed, Eq.~\eqref{energy norm} (with $\phi \sim \omega \psi =0$) gives a vanishing result,
\begin{equation}
    \|\phi_0 \|^2_{_E} =  \dfrac{r_h^2 \Lambda}{3\chi^3} \left[  1 - \chi^2 \left( 1 + \dfrac{r_\Lambda}{r_h}  + \dfrac{r_\Lambda^2}{r_h^2}      \right)       \right]\Bigg|_{\chi_h}^{\chi_\Lambda}
    = 0.
\end{equation}

\begin{figure}\centering
    \includegraphics[scale=0.665]{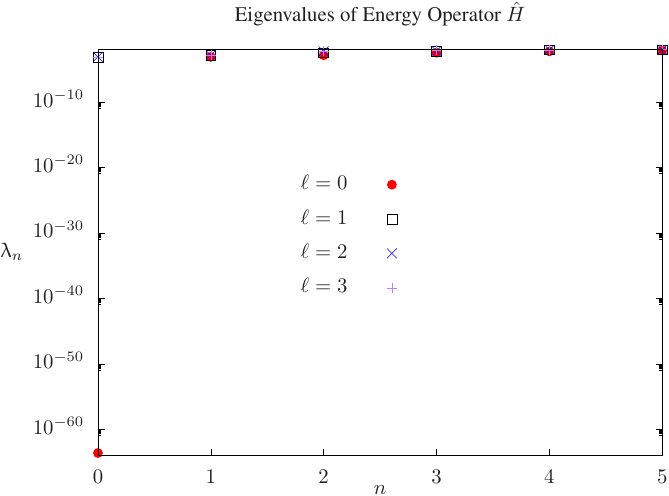}
    \caption{Eigenvalues of the discretized representation of the energy operator $\hat H$ (see Eqs.~\eqref{energy norm scalar} and \eqref{matrix evolution}) for scalar perturbations on a SdS spacetime with $\Lambda M^2 = 0.01$ and $N=100$ grid points. Scalar perturbations with $\ell=0$ (red circles) have a vanishing fundamental eigenvalue $\uplambda_0\sim 10^{-65}$ (up to round-off error) associated with the non-trivial zero mode. For $\ell > 0$, i.e. $\ell=1$ (black squares),  $\ell=2$ (blue `x' crosses) and $\ell=3$ (purple `+' crosses), $\hat H$ is positive-definite and $\uplambda_n >0$. Due to the scaling of the plot, all these modes are bunched together at the top.
    }
    \label{EnergyOperatoer_Eingevalue}
\end{figure}
%
\begin{figure}[t]\centering
    \includegraphics[scale=0.7]{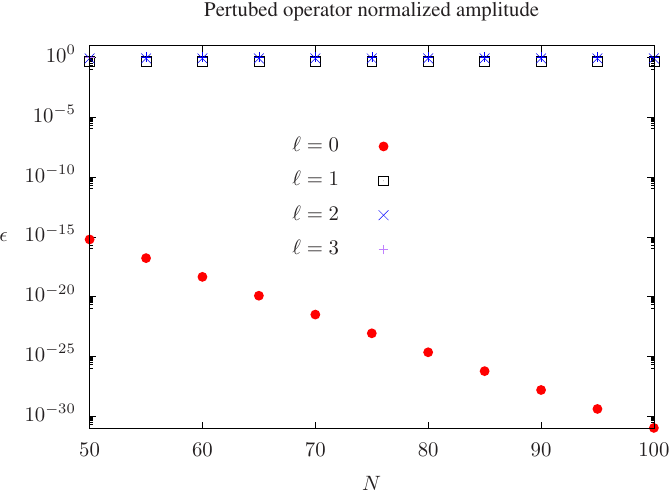}
    \caption{Numerical convergence of the normalized perturbation amplitude $\epsilon$ from Eq. \eqref{perturbed operator redefinition} for scalar fields on SdS spacetime with $\Lambda M^2 = 0.01$ and frequency $k=10$ for a deterministic perturbation added to the BH potential. For $\ell > 0$, i.e. $\ell=1$ (black squares), $\ell=2$ (blue `x' crosses) and $\ell=3$ (purple `+' crosses), $\|\delta L_1 \|_{_E}\sim 1$ and the normalized amplitudes converge to values with the same order of magnitude of the target scale (here $10^{-1}$). For $\ell=0$ (red circles), $\epsilon$ tends to zero as $N$ increases, due to the norm $\|\delta L_1 \|_{_E}$ being ill-defined. These extremely small values of the normalized perturbation amplitude explain why Ref.~\cite{Sarkar:2023rhp} did not observe the same spectral instabilities for $\ell=0$ as those for $\ell>0$.}
    \label{NormPertubedOperator}
\end{figure}

Consequently, the zero mode is also an eigenvector of the energy operator, with vanishing eigenvalue. 

We confirm this by calculating the eigenvalues of the discrete representation for the energy operator $\hat H$ (see Sec.~\ref{sec:NumDiscretization}). Fig.~\ref{EnergyOperatoer_Eingevalue} displays the eigenvalues of $\hat H$ for $\ell=0$ (red circles), $\ell=1$ (black squares),  $\ell=2$ (blue `x' crosses) and $\ell=3$ (purple `+' crosses). The cases with $\ell>0$ have real, positive eigenvalues $\uplambda_n$, as expected from a positive-definite operator. For $\ell=0$, we encounter $\uplambda_0 =0$ up to numerical round-off error, which in our case is $\uplambda_0\sim10^{-65}$ for $N=100$ grid points.

Formally, the existence of the zero eigenvalue prevents the energy operator $\hat H$ from being invertible, i.e. $\hat H$ is singular. In principle, this property prevents the calculation of the adjoint operator $L^\dagger$ via Eq.~\eqref{dagger operator}. In practice, numerical techniques compute finite large numbers for $\hat H^{-1}$ and $L^\dagger$ due to grid discretization with a finite $N$, in contrast to the exact analytical results. However, conclusions extracted from manipulating these discrete operators must be dealt with extreme care. 

Let us then consider the results from Ref.~\cite{Sarkar:2023rhp} on the spectral stability of scalar $\ell=0$ QNMs under perturbations of the SdS potentials. A direct assessment of the stability follows from calculating the QNMs of the operator with a small perturbation, as indicated in Eq.~\eqref{perturbed operator}. The particular type of perturbation $\delta L$, in accordance with the setup of Ref.~\cite{Jaramillo:2020tuu}, is one which only affect the effective potential $q(\chi)$ in Eq.~\eqref{eq:Operators_L1_L2},
\begin{equation}
\label{perturbed potential normalisation}
\delta L = \dfrac{\delta L_1}{ \|\delta L_1 \|_{_E}}, \quad 
\delta L_1 = 
\left(
	\begin{array}{cc}
		0 & 0 \\
		-\delta q(\chi)/w(\chi) & 0
	\end{array}
\right).
\end{equation}
Note that the factor $ \|\delta L_1 \|_{_E}$ in the above expression ensures that $\| \delta L \|_{_E} = 1$ as required by Eq.~\eqref{perturbed operator}. Effectively, one can redefine Eq.~\eqref{perturbed operator} in the form
\begin{equation}
\label{perturbed operator redefinition}
L \rightarrow L + \epsilon \, \delta L_1, \quad \epsilon \rightarrow \dfrac{\epsilon}{\|\delta L_1 \|_{_E}},
\end{equation}
where $\epsilon$ is redefined such that it represent an ``energy-normalized'' perturbation scale\footnote{From this point on, $\epsilon$ will refer to the normalized perturbation scale if not stated otherwise.}. For $\ell=0$, the normalized perturbation amplitude $\epsilon$ in Eq. \eqref{perturbed operator redefinition} is ill-defined, since the quantity $\|\delta L_1 \|_{_E}$ relies on the existence of the inverse operator $\hat H^{-1}$.

As a representative example, Fig.~\ref{NormPertubedOperator} shows the normalized perturbation scale $\epsilon$ from Eq. \eqref{perturbed operator redefinition} for the cases $\ell=0$ (red circles), $\ell=1$ (black squares),  $\ell=2$ (blue `x' crosses) and $\ell=3$ (purple `+' crosses) as a function of the numerical resolution $N$. Following Ref.~\cite{Sarkar:2023rhp}, we consider a deterministic perturbation
\beq 
\label{eq:deterministic pertubations}
\delta q(\chi) = \cos(2\pi k \chi),
\eeq
 with targeted amplitude $10^{-1}$. For the wavenumber, we set $k=10$, as opposed to $k=60$ employed in Ref.~\cite{Sarkar:2023rhp}. The reason is that an accurate discrete representation of $\delta q(\chi)$ for $k=60$ requires a rather high grid resolution ($N\sim 10^3$), and the relevant conclusions can be drawn with considerably less numerical expenditure. For $\ell >0$, one finds approximately $\|\delta L_1 \|_{_E}\sim 1$, and the normalized amplitude $\epsilon$ already has the same order of magnitude as the targeted value. Moreover, its numerical value converges to a fixed value as the numerical resolution $N$ increases. The $\ell = 0$ case, on the other hand, tends to zero when we increase the numerical resolution, since $\|\delta L_1 \|_{_E}$ is ill-defined. i.e. $\|\delta L_1 \|_{_E}\rightarrow \infty$ as $N\rightarrow \infty$.

Hence, one may be wrongly lead to the conclusions that the $\ell=0$ QNMs possess stronger spectral stability properties than the cases where $\ell>0$, if one uses this nearly-zero normalized amplitude $\epsilon$ and sees no significant displacement in the lower part of the QNM spectrum (the fundamental mode and first overtones). As we will show in Sec.~\ref{sec:RN-dS}, the $\ell=0$ spectra do suffer from the same instabilities as $\ell > 0$ if one promotes the normalized amplitude $\epsilon$ as a free parameter (not normalized by the energy norm) to quantify the perturbation's scale. This approach is appropriate due to the fact that the perturbation's targeted scale and its normalized scale $\epsilon$ have typically the same order of magnitude whenever $\|\delta L_1 \|_{_E}$ is well-defined, as we discussed above. We assume the validity of this property in a limiting approach $\ell \rightarrow 0$, and work directly with $\epsilon$ as a free (non-normalized) perturbation parameter when $\ell = 0$.

\begin{figure}[t]
\includegraphics[scale=0.343]{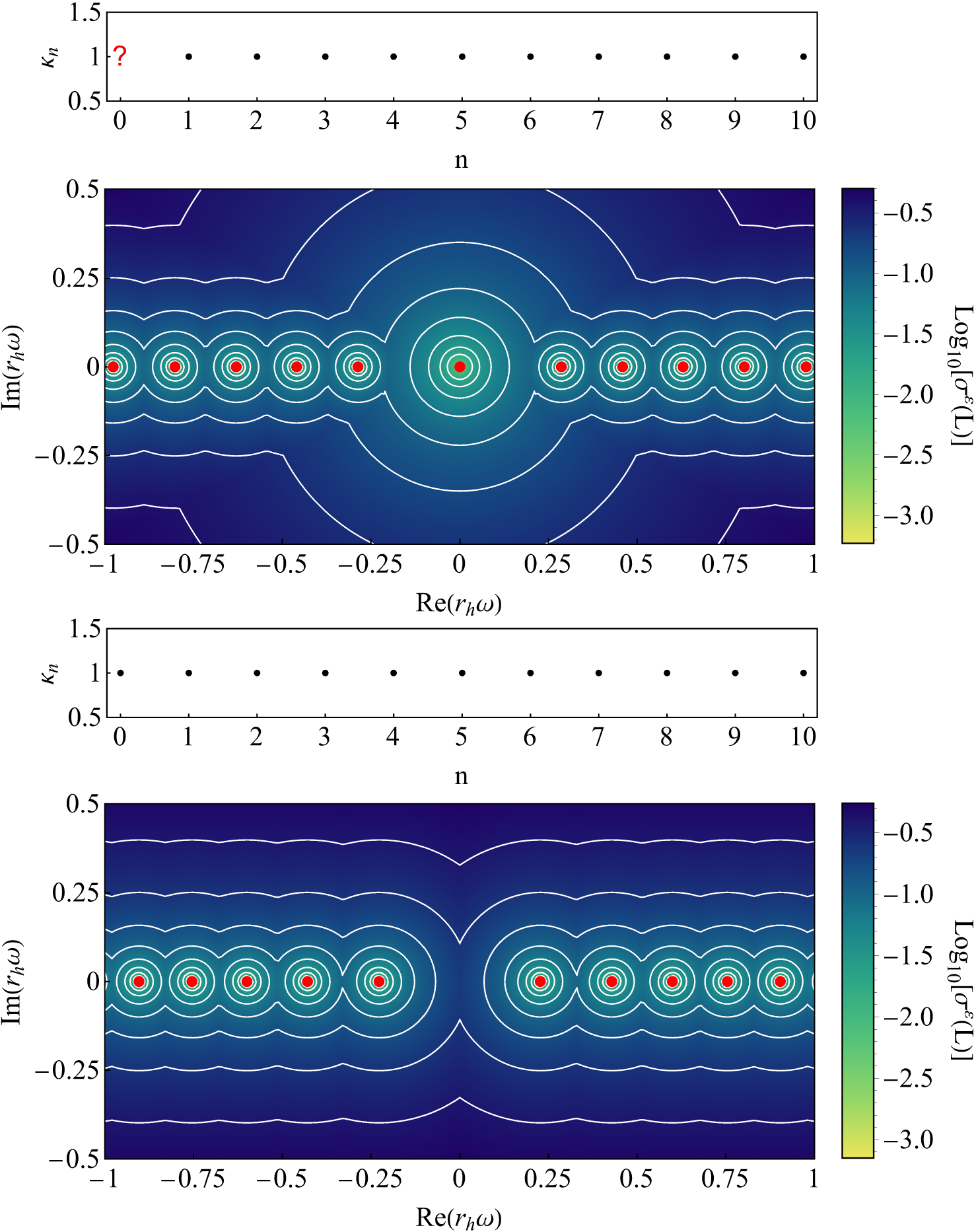}
\caption{Top panel: Scalar $\ell=0$ normal modes (red dots) and pseudospectra (white contour curves) of a SdS BH with $\Lambda M^2=0.01$. The operator $L$ in this test case is self-adjoint, since $L_2=0$. The $\epsilon$-pseudospectra contours range from $-1.8$ (innermost contour) to $-0.4$ (outermost contour) with steps of $0.2$. On top of figure, the condition number is shown for various modes. The question mark for the non-trivial zero mode corresponds to a divergent condition number as the grid points are increased. Bottom panel: Same as top panel for scalar $\ell=1$ normal modes. For both figures, $N=150$ grid points were used.}\label{self_adjoint}
\end{figure}

\begin{figure*}[]
\centering
\includegraphics[scale=0.343]{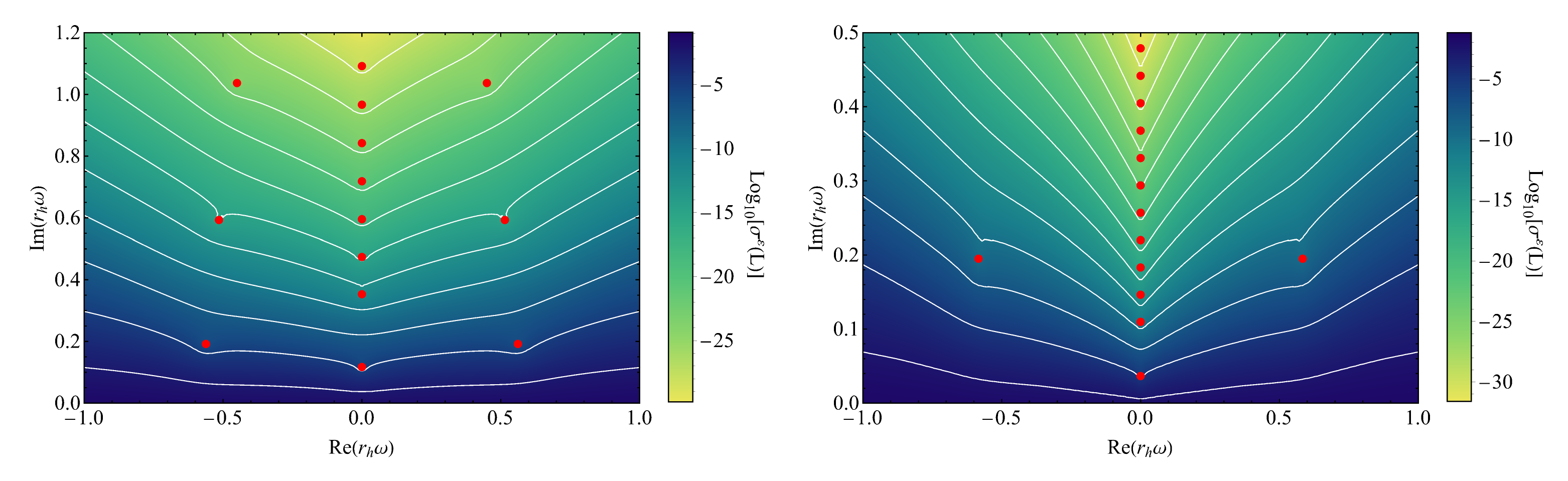}
\caption{Left: Scalar $\ell=1$ QNMs (red dots) and pseudospectra (white contour curves) of a SdS BH with $\Lambda M^2=0.01$. The $\epsilon$-pseudospectra contours range from $-27.5$ (uppermost contour) to $-2.5$ (bottom contour) with steps of $2.5$ and $N=150$. Right: Same as left with $\Lambda M^2=0.001$. The $\epsilon$-pseudospectra contours range from $-30$ (uppermost contour) to $-2.5$ (bottom contour) with steps of $2.5$ and $N=250$.}\label{SdS1}
\end{figure*}

\subsubsection{$L_2=0$: The self-adjoint case}\label{sec:l0_selfadjoint}
We have analytically and numerically established that calculations involving the energy norm associated with the scalar $\ell=0$ QNM problem of BHs in dS geometries are not to be taken at face value. However, the numerical discretization of the operators we use leads to an effective cutoff to the divergences involved. And while some quantities are manifestly unphysical even with this cutoff, such as the normalized perturbation amplitudes discussed above, others are seemingly better behaved. Particularly, a calculation of the pseudospectrum at a finite numerical resolution seems to yield a result which is similar to those for other values of $\ell$, both qualitatively and quantitatively.

Motivated by Ref. \cite{Jaramillo:2020tuu}, we perform a test to verify whether the numerically evaluated pseudospectrum provides a qualitatively correct picture by first performing a calculation for which we set $L_2=0$. This choice cancels the dissipative boundary conditions at the event and cosmological horizon, thus rendering the operator $L$ selfadjoint. The system becomes conservative and its eigenvalues become normal modes. Figure \ref{self_adjoint} shows two pseudospectra for a conservative SdS system for which we have set $L_2=0$. In the bottom panel, we observe a case of a scalar field with $\ell=1$. The pseudospectrum is well-behaved, as expected given the discussion in the previous sections. The boundaries of the $\epsilon$-pseudospectral regions form concentric circles of radius $\epsilon$ around the modes, indicating spectral stability. The condition numbers of the modes tell the same story; since $L$ is selfadjoint, the left and right eigenvectors of each eigenvalue coincide, leading to $\kappa_n=1$.

In the top panel of Fig. \ref{self_adjoint}, we show the pseudospectrum of the $L_2=0$ construction corresponding to an $\ell=0$ scalar field. The picture is similar to the $\ell=1$ case, though with one particular oddity. Although the concentric circles around modes $\omega \neq 0$ have radii of approximately $\epsilon$, the circles around the zero mode are about twice as large, modulating the function as they spread out. This indicates that, while the operator is still manifestly selfadjoint, the divergence of the resolvent around the zero mode has a greater ``width", suggesting that it captures not only the standard singularity present at an eigenvalue, but also a divergence of another nature, namely, the one due to the norm. What is perhaps surprising is that at small enough values of $\epsilon$ the effect seems to be contained to a vicinity of the zero mode itself, suggesting that the pseudospectrum still provides a qualitatively valid description elsewhere. We will make use of this fact in interpreting the $\ell=0$ pseudospectra we compute for the RNdS case in the next subsection. We leave this analysis for the RNdS case for the sake of brevity, and generality.

We also note that the special status of the $\omega=0$ mode is also observed when we calculate the condition number via Eq.~\eqref{condnumb}. Our results give $\kappa_n=1$ for all modes with $\omega \neq 0$, as expected (see inset in the top panel of Fig.~\ref{self_adjoint}). However, Eq.~\eqref{condnumb} is not directly applicable for $\omega=0$, because the zero mode has a vanishing energy norm, i.e., $\|u_n\|_{_E} = \|v_n\|_{_E} = |\langle u_n,v_n\rangle|_{_E} = 0$. Numerically, the ill-defined condition number manifests itself with values of $\kappa_0$ that tend to a divergence as the resolution $N$ increases (designated with a red question mark on the top panel of Fig. \ref{self_adjoint}). While formally it may be possible that $\kappa_0\rightarrow 1$ if one considers the limit $\ell \rightarrow 0$ in a suitable manner, developing this hypotheses goes beyond the scope of the present work.

\subsubsection{Schwarzschild-de Sitter pseudospectra}

For completeness, in Fig. \ref{SdS1} we present two plots of SdS scalar-field pseudospectra, corresponding to two representative examples in the parameter space. Particularly, we set the angular number $\ell=1$, and we use the values of the cosmological constant $\Lambda M^2=10^{-2}$ and $\Lambda M^2=10^{-3}$. We observe the expected picture of the spectral instability of QNMs~\cite{Jaramillo:2020tuu,Sarkar:2023rhp}, with contour branches that open up in the complex plane.

We note that besides the PS QNMs, purely-imaginary modes appear as well. These are not the numerically non-convergent ``branch-cut'' modes, such as those found in asymptotically-flat compact objects \cite{Jaramillo:2020tuu,Destounis:2021lum,Boyanov:2022ark}, but rather convergent eigenvalues of the QNM spectrum that belong to the dS family. In both cases depicted in Fig. \ref{SdS1}, the dS family is the dominant one, i.e. the least damped mode is purely imaginary. Particularly, for the $\Lambda M^2=10^{-2}$ case, the dominant dS mode is followed by a PS oscillatory mode, followed by two more dS modes, and so on and so forth. For the case $\Lambda M^2=10^{-3}$, the fundamental mode and first two overtones belong to the dS family, with the next overtone belonging to the PS family, after which nine dS overtones take over before a PS mode reappears (that lies outside the range of the complex plane that we have plotted).

\begin{figure*}[t]
\centering
\includegraphics[scale=0.343]{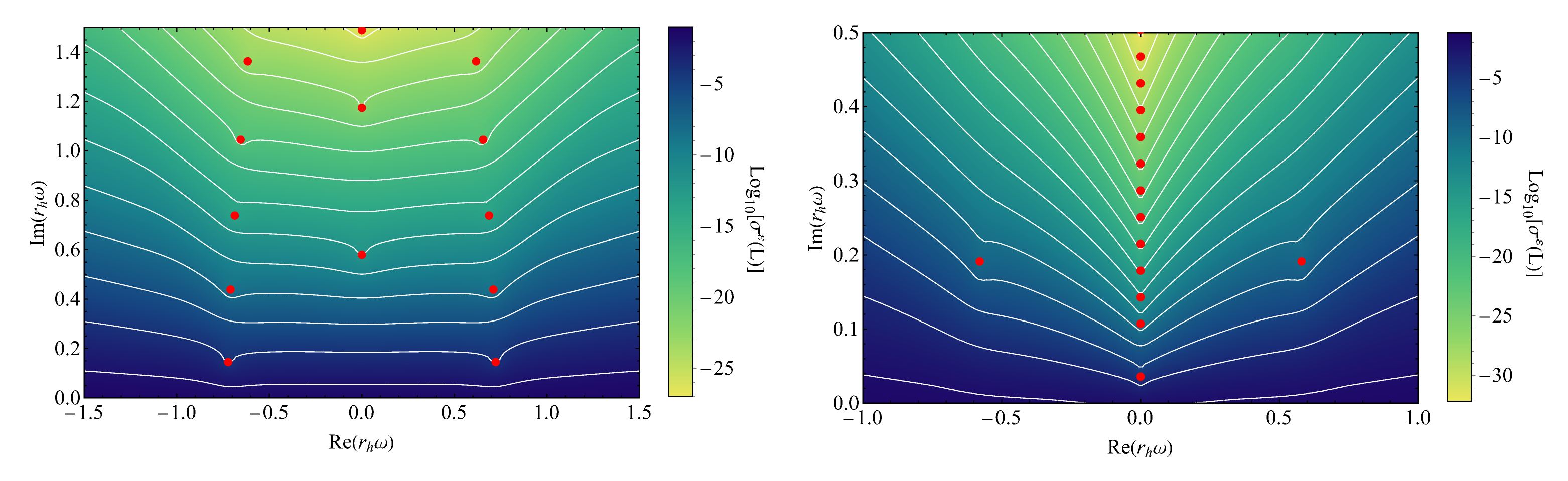}
\caption{Left: Scalar $\ell=2$ QNMs (red dots) and pseudospectra (white contour curves) of a RNdS BH with $\Lambda M^2=0.06$ and $Q/M=0.5$. The $\epsilon$-pseudospectra contours range from $-28$ (uppermost contour) to $0$ (bottom contour) with steps of $2$ and $N=150$. Right: Same as left for scalar $\ell=1$ QNMs of a RNdS BH with $\Lambda M^2=0.001$ and $Q/M=0.3$. The $\epsilon$-pseudospectra contours range from $-34$ (uppermost contour) to $0$ (bottom contour) with steps of $2$ and $N=250$.}\label{RNdS1}
\end{figure*}
%
\begin{figure*}[t]
\centering
\includegraphics[scale=0.343]{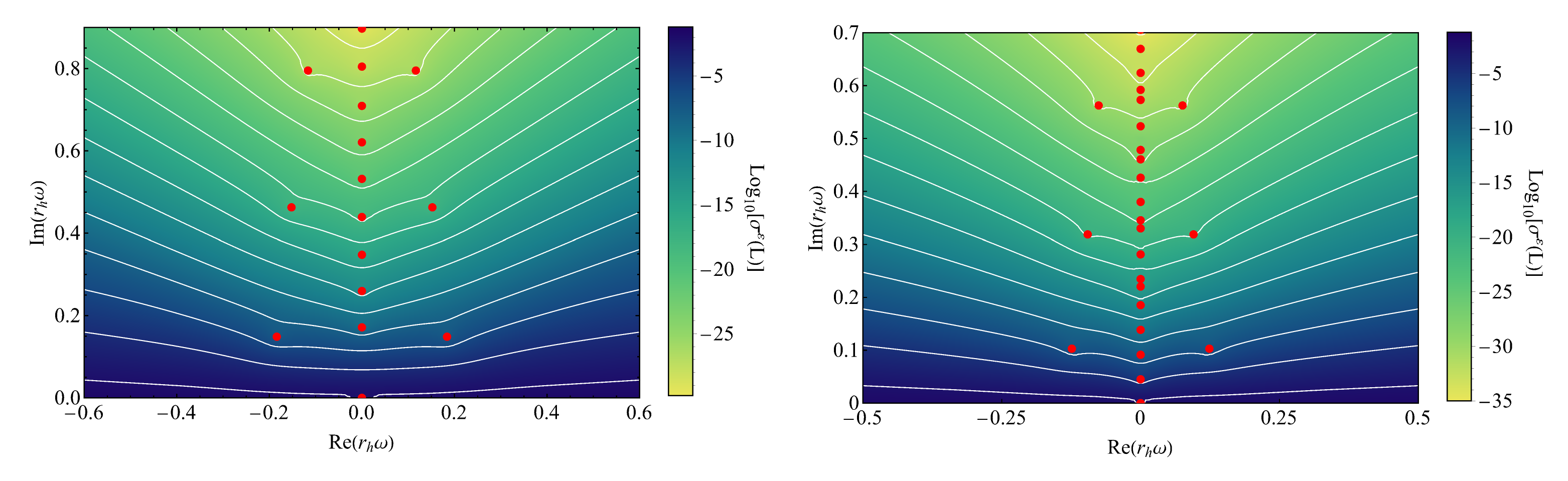}
\caption{Left: Scalar $\ell=0$ QNMs (red dots) and pseudospectra (white contour curves) of a RNdS BH with $\Lambda M^2=0.01$ and $Q/M=0.9$. The $\epsilon$-pseudospectra contours range from $-34$ (uppermost contour) to $0$ (bottom contour) with steps of $2$ and $N=150$. Right: Same as left panel with $\Lambda M^2=0.03$ and $Q/Q_{\rm{max}}=0.999$. The $\epsilon$-pseudospectra contours range from $-35$ (uppermost contour) to $-2.5$ (bottom contour) with steps of $2.5$ and $N=200$.}\label{RNdS2}
\end{figure*}

An important aspect of the pseudospectrum, both here and in the RNdS calculations presented below, is that the contour lines generally have their smallest imaginary part on the imaginary axis, from where they go up and presumably approach the logarithmic behavior described in~\cite{Destounis:2021lum}, but they are closer together in that region. This implies that the dS modes (as well as the NE ones in RNdS) are affected by lower-energy perturbations, but that their relative displacement may be less than that of PS modes when the latter are also destabilized. This may contribute to the likelihood of an ``overtaking" instability of the fundamental mode after a perturbation, as we will see in the next section for particular examples in the parameter space.

\subsection{Reissner-Nordstr\"om-de Sitter black holes}\label{sec:RN-dS}

\subsubsection{Reissner-Nordstr\"om-de Sitter pseudospectra}

We now move on to the RNdS case, presenting results involving the pseudospectrum and spectrum of a perturbed evolution operator, for a scalar field. As discussed above, as well as in \cite{Cardoso:2017soq}, RNdS geometries possess three families of QNMs: the complex PS modes, the purely imaginary dS, and purely imaginary NE modes. Since the NE modes are in the vicinity of dominant (longest-lived) modes only for RNdS geometries with charge which is quite close to the extremal one $Q_\text{max}$ (for which the inner and outer BH horizons almost overlap), and are particularly long-lived for the complicated case of $\ell=0$, it is interesting to test how the pseudospectra behave in this previously uncharted case.

Figure \ref{RNdS1} shows the pseudospectrum for two representative cases with different angular numbers $\ell\geq 1$, cosmological constants, and BH charges. In both cases the charge $Q$ is less than $50\%$ of the extremal, and $\ell\neq 0$, making the NE family of modes subdominant (in fact they lie outside the plotted region). On the one hand, we observe the effect of the cosmological constant and charge on the spectrum, which has been thoroughly analyzed in~\cite{Cardoso:2017soq,Cardoso:2018nvb,Destounis:2019hca}. Note that in the first case (on the left), the dominant mode belongs to the PS family, while the three purely imaginary modes that appear belong to the dS family, though on the right, the picture is similar to that of SdS BHs with $\Lambda M^2=10^{-3}$, where a plethora of dS modes dominate the spectrum in the part of the complex plane shown. On the other hand, we observe the standard QNM spectral instability, and we note that the minima of the contour lines once again generally lie on the imaginary axis, except for the lines in the vicinity of the fundamental mode and first overtone in the left panel, suggesting that this is a general feature for PS modes which are below any purely imaginary ones.

\begin{figure}[t]
\centering
\includegraphics[scale=0.39]{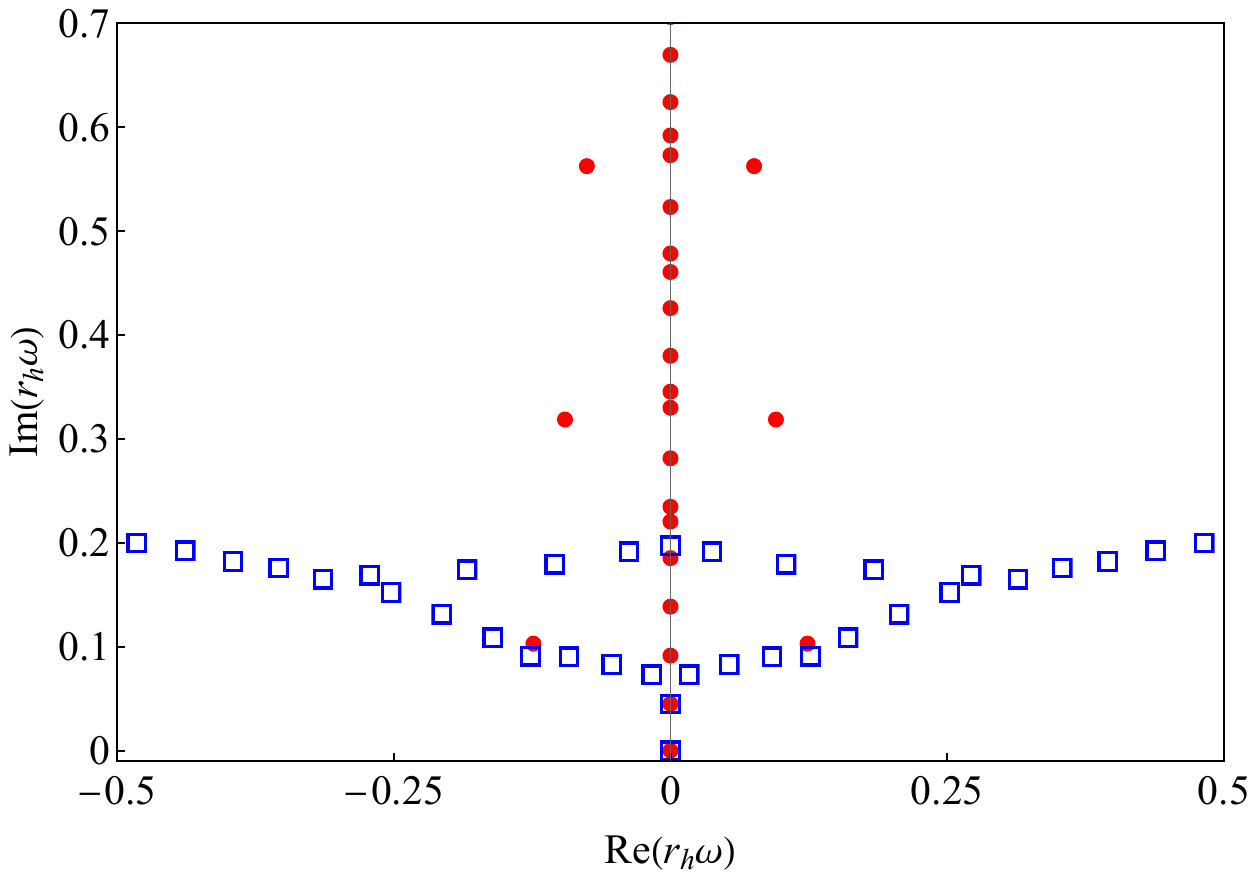}
\caption{Scalar unperturbed $\ell=0$ QNMs (red dots) and perturbed $\ell=0$ QNMs (blue boxes) of a RNdS BH with $\Lambda M^2=0.01$ and $Q/Q_{\rm{max}}=0.999$. The perturbed QNMs result from a random-vector perturbation added to the effective potential with non-normalized perturbation scale $\epsilon=10^{-3}$. The number of grid points used for both spectra is $N=200$.}\label{RNdSpert}
\end{figure}

In Fig. \ref{RNdS2} we show the pseudospectra of two RNdS BHs for $\ell=0$, for which the spectrum contains the non-trivial zero mode. As discussed in previous sections, this zero mode has an associated eigenfunction with a vanishing energy norm, which makes the calculation of operator norms generally ill-defined. However, as discussed above, qualitatively reasonable results may be obtained for the pseudospectrum for a given numerical resolution $N$, which acts as a cutoff to the divergences involved. Particularly, the results from Sec.~\ref{sec:l0_selfadjoint} provide a strong indication that the numerical cutoff may lead to meaningful results for the pseudospectrum in the vicinity of the overtones, for small values of $\epsilon$. Further evidence for this is given by the fact that the pseudospectral regions observed in Fig. \ref{RNdS2} have the trend that one expects, given the destabilization of the spectrum under perturbations, shown in Fig. \ref{RNdSpert}. We note that a qualitative picture of the pseudospectrum is all we can expect to obtain in any case, even when $\ell\ge1$, due to the general convergence issues discussed in Ref.~\cite{Boyanov:2023}. Therefore, for the particular calculation of the pseudospectrum from the finite-rank approximation to the resolvent operator and its norm, the result obtained for $l=0$ with the energy norm seems to be as good a qualitative description as one can expect at this level. For a more in-depth discussion on the issue of convergence, we again refer the reader to Ref.~\cite{Boyanov:2023}.

The left panel of Fig. \ref{RNdS2} depicts the pseudospectrum associated with the results shown in Fig.~13 of Ref.~\cite{Sarkar:2023rhp} ($Q/M=0.9$ and $\Lambda M^2=10^{-2}$), for which a stronger spectral stability was suspected. However, the qualitative aspects of the pseudospectrum suggest the opposite, which is in line with our discussion in Sec.~\ref{sec:zero mode}. The opening of branches indicate a spectral instability of both complex PS and purely imaginary dS modes. As a side note, it is worth observing that in this left panel, the purely imaginary modes (which here belong to the dS family) seem almost equidistant as the overtone number increases, if one excludes the zero mode, in contrast with the figure in the right panel.

The right panel of Fig. \ref{RNdS2} depicts a case of a near-extremally-charged RNdS BH ($Q/Q_\text{max}=0.999$ and $\Lambda\, M^2 = 0.03$). In this case the part of the spectrum depicted contains all three families of QNMs: oscillatory PS modes, dS modes, and NE modes. Notice the interplay occurring in the imaginary axis, where now the equidistant dS modes are scrambled with NE modes, though both families can be clearly tracked with the use of the approximations provided in \cite{Cardoso:2017soq}. For the specified BH parameters, the NE family possesses the least-damped fundamental mode that dominates the late-time tail until the zero mode takes over to relax the ringdown signal to a non-zero constant, as discussed in \cite{Zhu:2014sya}. Again, the pseudosepctral contour lines indicate a spectral instability of all families of QNMs, as in all other cases studied here.

The presence of spectral instability in all cases studied in this work reflects an instability to physical perturbations. While the pseudospectrum ideally captures the possible migration of modes under any type of perturbation to the evolution operator $L$, perturbing only the effective potential is sufficient to trigger the instability. In Fig.~\ref{RNdSpert} we show explicitly that this also applies to the $\ell=0$ case. We plot an unperturbed QNM spectrum, along with the spectrum after a perturbation to the effective potential in the form of the addition of a random vector with components of magnitude $10^{-3}$ to its discretized version. For other values of $\ell$, such a vector perturbation has a corresponding energy norm of the same order, $10^{-3}$, suggesting that this is likely a very small perturbation in any physically reasonable, well-defined norm. In other words, we treat $\epsilon$ as a free parameter, extrapolating the likely order of magnitude of the perturbation from other known cases. We observe that for such a perturbation, overtones are indeed destabilized, independently of the family they belong to.

\subsection{Fundamental mode (in)stability}\label{sec:destabilization}

In this final subsection, we revisit another open issue put forward in Ref.~\cite{Sarkar:2023rhp} regarding the spectral stability of the fundamental QNM with respect to (deterministic) perturbations to the effective potential. In what follows, we will present quantitative results which show that the mode which is initially the fundamental one always appears to be spectrally stable, i.e. its relative displacement in the complex plane for perturbations of up to order $\epsilon=10^{-1}$ (in the energy norm) is less than $\epsilon$, in agreement with \cite{Sarkar:2023rhp}. However, we note that in certain regions of the parameter space, for which the imaginary part of a PS mode frequency is very close to one of the purely imaginary modes, and these two are the least damped modes, then an ``overtaking" instability may occur, whereby the mode which is initially the fundamental one becomes an overtone after the perturbation. Particularly, we observe this for fine-tuned cases of the parameter space in which the fundamental mode is initially of the dS family, and gets overtaken by the first overtone, which is of the PS family.

More precisely, following Refs.~\cite{Cheung:2021bol,Courty:2023rxk}, we will use $\Delta\omega^{(\epsilon)}\equiv |\omega^{(0)}_0-\omega_0^{(\epsilon)}|/|\omega^{(0)}_0|$ to quantify the relative migration distance between the unperturbed fundamental mode $\omega_0^{(0)}$ and the perturbed fundamental mode $\omega_0^{(\epsilon)}$. In this definition, the fundamental mode is understood as the mode with the smallest imaginary part. Ref.~\cite{Cheung:2021bol} identifies two types of spectral instability for the fundamental mode: (i) a destabilization via a continuous migration of the initial fundamental mode itself for which $\Delta\omega^{(\epsilon)} \gg \epsilon$; (ii) a discontinuous overtaking of the fundamental mode, for which $\Delta\omega^{(\epsilon)} \gg \epsilon$ due to a migrated overtone becoming the new fundamental mode. This latter case is the one observed here.

We study deterministic perturbations to the potential of the form \eqref{eq:deterministic pertubations} and focus on perturbations with wavenumber $k=20$, which is at the high end of what is smoothly resolved with our number of discretization points (we use wavenumber perturbations because these are the ones which destabilize the QNM spectrum the most, as shown in \cite{Jaramillo:2020tuu}). We use a normalized perturbation scale ranging in $\epsilon=10^{-6}-10^{-1}$.
For perturbations of a different variety to the same class of BHs considered here, we refer the reader to Ref.~\cite{Courty:2023rxk}.

\begin{figure}[t]
\centering
\includegraphics[scale=0.4]{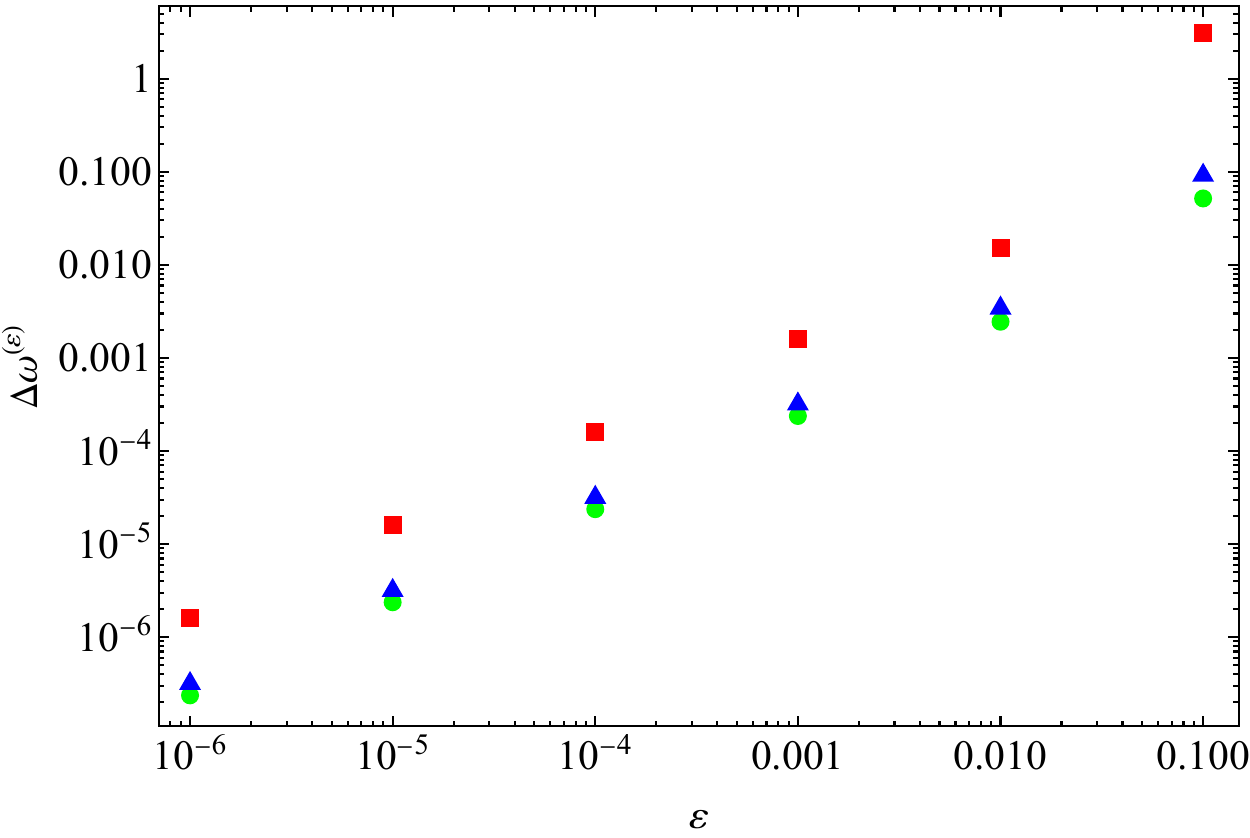}
\caption{Relative value of distance in the complex plane $\Delta \omega ^{(\epsilon)}$ between the unperturbed fundamental QNM and the perturbed fundamental mode versus the scale $\epsilon$ of the deterministic perturbation-vector with wavenumber $k=20$ that is added to the effective potential of various RNdS BHs. The cases depicted correspond to $\ell=1$, $Q/M=0.5$, and $\Lambda M^2=3/121\sim 0.025$ (red squares, overtaking instability from dS to PS family), $\ell=2$, $Q/M=0.5$ and $\Lambda M^2=0.06$ (green circles, PS dominant family) and, finally, $\ell=2$, $Q/M=0.5$ and $\Lambda M^2=0.006$ (blue triangles, dS dominant family). The number of grid points used for all cases is $N=150$ which guarantee an accurate resolution for the deterministic perturbation-vector.
}\label{RNdSpertstab}
\end{figure}

Figure \ref{RNdSpertstab} quantifies the migration of the fundamental QNM $\Delta\omega^{(\epsilon)}$ caused by this effective potential perturbation. We depict three RNdS cases with fixed charge $Q/M=0.5$ and varying $\Lambda M^2$, as well as $\ell$. We observe that when the BH parameters are such that the fundamental mode is an oscillatory PS one (green circles in Fig. \ref{RNdSpertstab}) or a purely imaginary dS one (blue triangles in Fig. \ref{RNdSpertstab}), and the first overtone has a sufficiently distinct imaginary part (as occurs in both of the aforementioned), then the fundamental mode after the perturbation remains in the same family. In these cases, we always encounter $\Delta\omega^{(\epsilon)}\lesssim \epsilon$, thus the fundamental mode remains spectrally stable.

The third case we analyze is a fine-tuned one for which the original fundamental mode belongs to the dS family, and the first overtone is of the PS family, with an imaginary part which is only $~10^{-3}$ larger (in units of $M$). The parameters which lead to this configuration are $\ell = 1$,  $Q/M=0.5$, and $\Lambda M^2=3/121\sim 0.025$. We observe that the fundamental mode is stable until a perturbation of $\epsilon=10^{-1}$, for which the overtaking occurs (red boxes in Fig. \ref{RNdSpertstab}). Since the new fundamental mode is at a distance which is over one order of magnitude larger than the perturbation scale of $10^{-1}$, an overtaking instability can be said to take place. The cause for this effect, in this case, seems to be related to the fact that the PS family has the tendency for larger migration when perturbed, as shown in \cite{Courty:2023rxk}, and discussed above in regards to the shape of the pseudospectrum contours.

\section{Conclusions}

We have systematically analyzed several key features of the QNM instability of (neutral) scalar perturbations on the SdS and RNdS spacetimes, which have proven to be an important testing ground of the phenomenon due to their richer spectrum as compared to asymptotically flat BHs~\cite{Cardoso:2017soq,Destounis:2019hca}. Following Ref.~\cite{Jaramillo:2020tuu}, we have expressed the QNM problem in the hyperboloidal framework, which imposes the appropriate boundary conditions geometrically, and we have studied the behavior of the spectrum under perturbations, and calculated the associated pseudospectrum, in several representative examples. Our analysis has mainly been focused on three points: (i) revisiting the case of the peculiar spectral stability of QNM overtones for $\ell=0$ scalar perturbations found in Ref.~\cite{Sarkar:2023rhp}, (ii) the (in)stability of the fundamental mode for other angular numbers $\ell\geq 1$, also discussed in \cite{Sarkar:2023rhp}, and (iii) the calculation of the pseudospectrum of RNdS scalar perturbations, which had not been explored in the literature yet.

Regarding the stability of overtones for $\ell=0$, we have found that it is simply an artefact of an ill-defined norm. Particularly, the $\ell=0$ scalar QNMs in dS BHs possess a non-trivial zero mode, which turns out to have a vanishing energy norm. This implies that the norm operator acting on the space of solutions has a zero eigenvalue, associated with this mode, and it is thus not invertible. Calculating norms numerically with this operator thus leads to rapidly diverging results, and attempting to normalize perturbations with such a norm leads to practically vanishing results. It is thus likely that such an ill-defined normalization has lead to a nearly vanishing perturbation to be dubbed ``large", and the overtones within the range of study to be mistaken for stable. In our analysis, by treating the perturbation scale as a non-normalized free parameter, we have shown that the $\ell=0$ QNMs are likely to be spectrally unstable, just like in all other cases studied in the literature thus far.

We have further explored the energy norm for $\ell=0$, finding that the cutoff induced by the spatial discretization can lead to qualitatively reasonable results in the calculation of some quantities, the pseudospectrum in the vicinity of the overtones (for small $\epsilon$) being one of them. From it, we also obtain a qualitative picture of spectral instability, both in SdS and RNdS BHs. We note that while we the qualitative description is a robust result, quantitative results on the convergence of the pseudospectrum are yet to be formalized due to the intricate notion of a Hilbert space for the QNM eigenfunctions~\cite{Warnick:2013hba,Gajic:2019oem,Gajic:2019qdd,galkowski2021outgoing}. An exploration of alternative norms in which the present issue may be resolved, along with the more general convergence problems of the pseudospectrum~\cite{Boyanov:2023}, is left for future work.

We have also examined the pseudospectrum of RNdS for cases with $\ell\geq 1$, for which the energy norm is positive-definite in the space of eigenfunctions, finding qualitatively the same picture of spectral instability in all of them. Specifically, we have looked at representative examples in both weakly-charged and near extremal cases, for which different families of modes are dominant.

Finally, the stability of the fundamental mode under high-wavenumber deterministic perturbations to the effective potential (akin to those used in~\cite{Jaramillo:2020tuu}) has been tested, particularly for cases $\ell\geq 1$, for which the magnitude of the perturbations can be measured well. We find that when the PS or dS family is sufficiently dominant with respect to the other families, the fundamental mode is stable. However, when the fundamental mode is of one family and the first overtone of another, and they have sufficiently similar imaginary parts, the migrated overtone can take over as the new fundamental mode after a sufficiently large perturbation. We have particularly observed this for a case in which a PS overtone overtakes a dS purely imaginary fundamental mode, and have dubbed this an ``overtaking" instability. We note, however, that the individual migration of each of the two modes in this case was within the bounds of spectral stability.

Ultimately, we have taken another step towards confirming the universality of QNM spectral instabilities of BHs and compact objects in spherical symmetry. These may become relevant for the BH spectroscopy program when the signal-to-noise ratio of current and future ground- and space-based detectors is increased sufficiently, giving us detections of ringdown signals which go beyond the dominant early-time emission~\cite{Jaramillo:2021tmt,Berti:2022xfj}.

\begin{acknowledgments}
The authors would like to warmly thank Vitor Cardoso and Jos\'e-Luis Jaramillo for stimulating discussions and comments.
K.D. acknowledges financial support provided under the European Union’s H2020 ERC, Starting Grant agreement no. DarkGRA–757480 and the MIUR PRIN and FARE programmes (GW-NEXT, CUP:B84I20000100001). K.D. also acknowledges hospitality and financial support from the Center for Astrophysics and Gravitation (CENTRA) where part of this work was conducted.
R.P.M acknowledges support from the Villum Investigator program supported by the VILLUM Foundation (grant no. VIL37766) and the DNRF Chair program (grant no. DNRF162) by the Danish National Research Foundation. This project has received funding from the European Union's Horizon 2020 research and innovation programme under the Marie Sklodowska-Curie grant agreement No 101131233
V.B. acknowledges support form the Spanish government projects PID2020-118159GB-C43 and PID-2020-118159GB-C44.
This project has received funding from the European Union's Horizon 2020 research and innovation programme under the Marie Sklodowska-Curie grant agreement No 101007855.
We thank FCT for financial support through Project~No.~UIDB/00099/2020.
We acknowledge financial support provided by FCT/Portugal through grants PTDC/MAT-APL/30043/2017 and PTDC/FIS-AST/7002/2020.
This research project was conducted using the computational resources of ``Baltasar Sete-Sois'' cluster at Instituto Superior T\'ecnico.
\end{acknowledgments}

\bibliography{dS_biblio}

\end{document}